\documentclass{elsart}

\usepackage{graphics}
\usepackage{graphicx}

\usepackage{amssymb,amsmath}
\usepackage[latin1]{inputenc}
\newcommand{\mgb}{MgB$_2$}
\begin{document}

\begin{frontmatter}



\title{Fluxon dynamics in Li-Al codoped \mgb\ by microwave surface resistance measurements}

\author[label1,cor1]{M. Bonura},
\author[label1]{A. Agliolo Gallitto},
\author[label1]{M. Li Vigni},
\author[label2]{M. Monni},
\address[label1]{CNISM and Dipartimento di Scienze Fisiche ed Astronomiche,
Universit\`{a} di Palermo, via Archirafi 36, I-90123 Palermo, Italy}
\address[label2]{SLACS-INFM/CNR and Dipartimento di Scienze Fisiche, Università degli Studi di Cagliari, I-09124 Monserrato (CA), Italy}
\corauth[cor1]{Tel.: +39 0916234261; fax: +39 0916162461; e-mail:
marco.bonura@fisica.unipa.it}
\begin{abstract}
The magnetic-field-induced variations of the microwave surface resistance, $R_s$, have been investigated in ceramic Mg$_{1-x}$(LiAl)$_x$B$_2$, with $x$ in the range $0.1 \div 0.4$. The measurements have been performed on increasing and decreasing the DC magnetic field, $H_0$, at fixed temperatures. At low temperatures, we have observed a magnetic hysteresis in the $R_s(H_0)$ curves in all the investigated samples. On increasing the temperature, the range of $H_0$ in which the hysteretic behavior is visible shrinks; however, in the sample with $x=0.1$ it is present up to temperatures close to $T_c$. We show that the field dependence of $R_s$ can be quantitatively justified taking into account the critical-state effects on the fluxon lattice only in the sample with $x=0.4$. On the contrary, in the samples with $x < 0.4$ the hysteresis exhibits an unusual shape, similar to that observed in others two-gap MgB$_2$ samples, which cannot be justified in the framework of the critical-state models.
\end{abstract}


\end{frontmatter}

\section{Introduction}
It has already been shown that the in-field microwave (mw) response of two-gap MgB$_2$ superconductor cannot be accounted for by standard models for fluxon dynamics~\cite{shibata,Sarti,noi-irrad}. The anomalies have been ascribed to the unusual structure of fluxons, related to the different field dependence of the two superconducting gaps~\cite{eskil,koshelev}. Support to this hypothesis has been given from investigation of a strongly neutron-irradiated \mgb\ sample~\cite{noiDEPIN}, in which the two gaps merged into a single value, and in a two-gap sample only at applied magnetic fields higher than the field at which the smaller gap is almost suppressed~\cite{noiGIANNI} (i.e. when fluxons assume a more conventional shape).

According to the theory of multi-band superconductivity, the inclusion of defects, by chemical substitution or irradiation, would increase the inter-band scattering and, consequently, change the relative magnitude of the two gaps. Investigation carried out in neutron irradiated samples has shown that high-level irradiation produces the merging of the two gaps~\cite{putti2,gonnelli2}. Recently, samples of Mg$_{1-x}$(LiAl)$_x$B$_2$, with $x$ in the range $0.1\div 0.4$, have been investigated by point-contact Andreev-reflection spectroscopy~\cite{dagheroAlLi}. A clear evidence has been given that the samples with $x=0.1$ and $x=0.2$ have a typical two-gap structure, while in the sample with $x=0.4$ the two gaps merge into a single value. On the contrary, in the sample with $x=0.3$ one or two gaps are apparently observed depending on the local critical temperature of the junction. In this paper, we report results of the magnetic-field dependence of the mw surface resistance, $R_s$, in the same samples investigated in Ref.~\cite{dagheroAlLi}. We have found that only the results obtained in the sample with $x= 0.4$ can be accounted for by using a model of fluxon dynamics in single-gap superconductors. In the other samples, we have observed a magnetic hysteresis of unusual shape, similar to that observed in other two-gap \mgb\ samples, which cannot be justified in the framework of the critical-state models. Our results confirm that mw measurements may discriminate the gap-structure of \mgb\ samples.

\section{Experimental apparatus and samples}\label{sec:samples}
The mw surface resistance has been investigated in ceramic samples of Mg$_{1-x}$(LiAl)$_x$B$_2$, with $x=0.1 \div 0.4$, in the range of temperatures $3 \div 40~\mathrm{K}$ and DC magnetic fields up to 1~T. The procedure for the preparation of the samples and their properties are reported in Refs.~\cite{dagheroAlLi,monniLiAl}. It is worth remarking that structural characterization by neutron and x-ray diffraction has highlighted that Li enters the \mgb\ structure in an amount less than the nominal one~\cite{monniLiAl}; in particular, the correct stoichiometry is Mg$_{1-x}(\mathrm{Al}_{\alpha} \mathrm{Li}_{1-\alpha})_x \mathrm{B}_2$, with $\alpha \sim 2/3$. The measured $T_c$ of the samples decreases with $x$, from 36~K down to 20~K, however it depends on the used method of measure; the values of the transition width increase with $x$ from 1.5~K up to 4.4~K.

The mw surface resistance has been measured by the cavity-perturbation technique. A copper cavity, of cylindrical shape with golden-plated walls, is tuned in the $\mathrm{TE}_{011}$ mode resonating at 9.6~GHz; the sample is located in the center of the cavity, where the mw magnetic field is maximum. The cavity is placed between the poles of an electromagnet which generates DC magnetic fields up to $\mu_0 H_0=1$~T. Two additional coils, independently fed, allow compensating the residual field and working at low magnetic fields. Details of the experimental apparatus and the field geometry are reported in Ref.~\cite{noi-irrad}. In all the measurements, $\emph{\textbf{H}}_{0}$, is perpendicular to the mw magnetic field, $\emph{\textbf{H}}_{\omega}$. This implies that, when the sample is in the mixed state, the induced mw current causes a tilt motion of the vortex lattice~\cite{BRANDT}. In this field geometry, the effects of the critical state of the fluxon lattice in the field dependence of $R_s$ are particular enhanced, because one detects the response of the whole fluxon lattice, and a magnetic hysteresis is expected~\cite{noistatocritico,noiisteresi}.

\section{Experimental results and discussion}\label{experimental}
\begin{figure}[th]
\centering \includegraphics[width=7.5 cm]{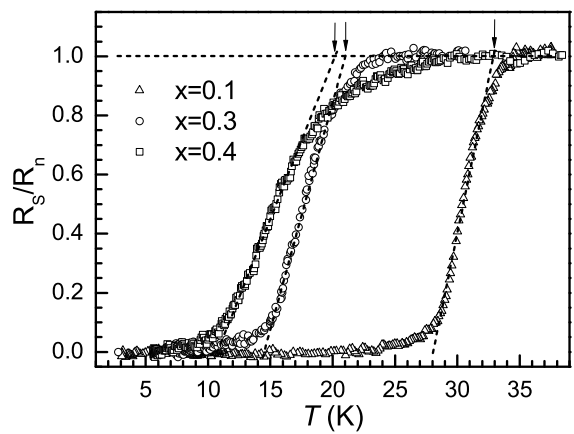}
\caption{Temperature dependence of the mw surface resistance ($R_n$ is the normal-state value), obtained for Mg$_{1-x}$(LiAl)$_x$B$_2$, with different LiAl content, at $H_0=0$. It is shown the procedure followed to determine $T_c$.}\label{Rs(T)}
\end{figure}

Fig.~\ref{Rs(T)} shows $R_s$, normalized to its normal-state value, $R_n$, as a function of the temperature for the samples with $x=0.1$, $x=0.3$ and $x=0.4$ at $H_0=0$. On increasing the LiAl content the transition broadens and shifts toward lower temperatures. Because of the wide transition and the large uncertainty of the values reported in the literature for $T_c$, in this paper we use for $T_c$ the values obtained by the method shown in the figure, which gives $T_c\approx 33.1\pm 0.2$~K for $x=0.1$, $T_c\approx 20.9\pm 0.2$~K for $x=0.3$ and $T_c\approx 20.2\pm 0.5$~K for $x=0.4$.

\begin{figure}[t]
\centering \includegraphics[width=7.5 cm]{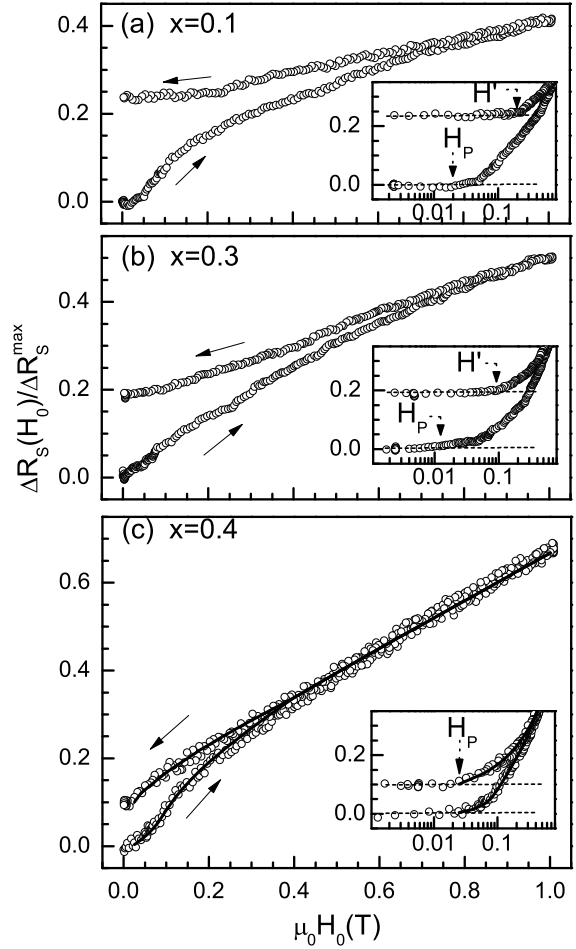}
\caption{Field-induced variations of $R_s$ at $T/T_c\approx 0.2$, obtained for Mg$_{1-x}$(LiAl)$_x$B$_2$, with different LiAl content. The insets show the curves in a logarithmic scale.}\label{fig1}
\end{figure}

The field-induced variations of $R_s$ have been measured at different temperatures. For each measurement, the sample was ZFC down to the desired temperature; the DC magnetic field was increased from 0 up to 1~T and, successively, decreased down to zero. Fig.~\ref{fig1} shows the field-induced variations of $R_s$, obtained at $T/T_c\approx0.2$. $\Delta R_s(H_0)\equiv R_s(H_0,T)-R_{res}$, where $R_{res}$ is the residual mw surface resistance at $T=3$~K and $H_{0}=0$; moreover, the data are normalized to the maximum variation, $\Delta R_s^{max}\equiv R_{n}-R_{res}$. The insets show the curves in a logarithmic scale to better highlight the low-field behavior; the arrow in the increasing-field branch identifies the first-penetration field of fluxons, $H_p$; the arrow in the decreasing-field branch indicates a characteristic field, $H^{\prime}$, below which $R_s$ does not change anymore.

In all the samples, the $R_s(H_0)$ curve exhibits a hysteresis, which is ascribable to the different magnetic induction, $B$, at increasing and decreasing $H_0$, due to the critical state of the vortex lattice~\cite{noiisteresi}; however, some differences have to be remarked. By comparing the decreasing-field branch of the $R_s(H_0)$ curves of the samples with $x=0.1$ and $x=0.4$, one can see that for $x=0.4$, $R_s$ monotonically decreases down to field values comparable with $H_p$; this behavior is expected and it is related to the monotonic decrease of $B$ when $H_0$ is decreased from the maximum value reached down to $H_p$. On the contrary, for $x=0.1$, the decreasing-field branch of the $R_s(H_0)$ curve exhibits a plateau starting from $\mu_0 H_0\approx 0.2$~T, although this value is about 10 times larger than $H_p$. We wish to remark that we have observed a similar behavior in all the two-gap \mgb\ samples we have investigated~\cite{EUCAS2007}. The presence of this plateau cannot be justified in the framework of the critical-state models, because it should indicate that the trapped flux does not change anymore on decreasing the applied field below 0.2~T.

In the sample with $x=0.3$, a plateau in the decreasing-field branch of the $R_s(H_0)$ curve is also present, but in a field range smaller than that characterizing the decreasing-field branch of Fig.~\ref{fig1}a. According to the results of Ref.~\cite{dagheroAlLi}, it could be due to the fact that the LiAl content is not homogenously distributed over the sample; consequently, the sample contains regions with characteristic two-gap superconductivity and regions at single gap.

In Refs.~\cite{noistatocritico,noiisteresi}, we have investigated the effects of the critical state of the fluxon lattice on $R_s(H_0)$ and have elaborated a model, in the framework of the Coffey and Clem theory~\cite{CC}, which quantitatively describes the hysteretic behavior of $R_s(H_0)$. The model has been elaborated considering that, because of the non uniform distribution of $B$, different regions of the sample contribute differently to the field-induced energy losses. So, in order to calculate $R_s(H_0)$, one has to determine the $B$ profile inside the sample, related to the field dependence of the critical current density, $J_c(B)$, and calculate a proper averaged value of $R_s(B)$ over the whole sample. From this model, it is expected that the shape of the hysteresis is strongly related to that of the magnetization curve, giving rise to a higher value of $R_s$ at decreasing fields than at increasing fields and a monotonic decrease of $R_s$ when the field is decreased from the maximum value reached down to $H_p$. The results obtained in the sample with $x=0.4$ have been analyzed using just this model~\cite{noiisteresi}. The parameters necessary to fit the data are the upper critical field, $H_{c2}$, $J_c(B)$, the depinning frequency~\cite{gittle}, $\omega_0$, and its field dependence. The line in Fig.~\ref{fig1} is the best-fit curve; it has been obtained using $\mu_0 H_{c2}=1.6$~T; $J_c(B)=J_{c0}(1-\beta B)$, with $J_{c0}=9\times 10^7~\mathrm{A/m}^2$ and $\beta=2.5~\mathrm{T}^{-1}$; $\omega_0(B)=\omega_{00}(1-\gamma B)$, with $\omega_{00}=6.2~\mathrm{GHz}$ and $\gamma=0.625~\mathrm{T}^{-1}$.

\begin{figure}[b]
\centering \includegraphics[width=7.5 cm]{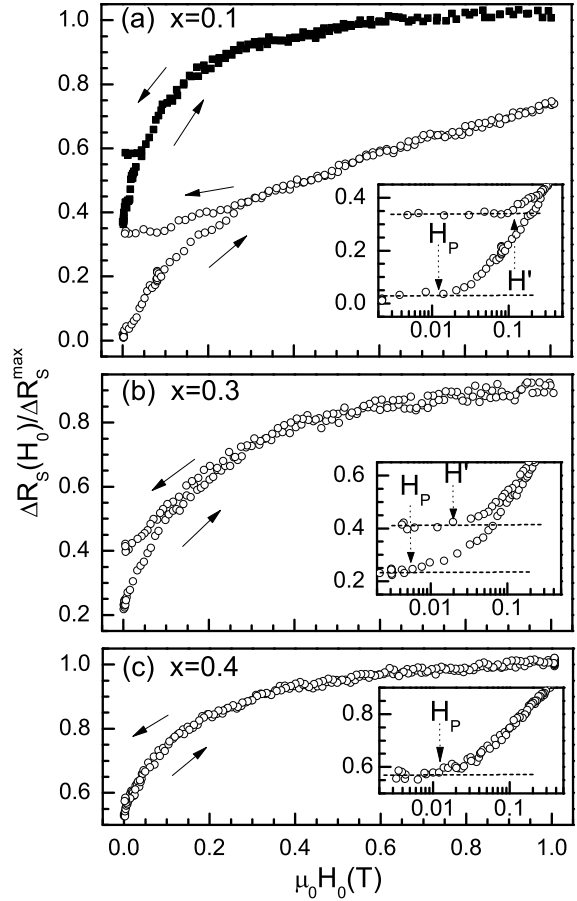}
\caption{Field-induced variations of $R_s$ for the Mg$_{1-x}$(LiAl)$_x$B$_2$ samples at temperature near $T_c$. Open symbols refer to results obtained at $T/T_c\approx 0.75$; full symbols in panel (a) are the results obtained at $T/T_c \approx 0.94$.}\label{nearTc}
\end{figure}

In Fig.~\ref{nearTc} we show a comparison among the $R_s(H_0)$ curves obtained in the different samples at $T/T_c\approx 0.75$ (open symbols); one more curve at $T/T_c \approx 0.94$ is shown in panel (a) for $x=0.1$ (full symbols). Comparing the curves obtained for different $x$ at $T/T_c\approx 0.75$, one can see that in the sample with $x=0.4$ no hysteresis is detectable, most likely because of the weakness of the pinning at this temperature. In the other samples, the hysteresis with the characteristic plateau is still present, even if it is visible in a restricted field range. The curve represented in panel (a) by full squares shows that in the sample with $x=0.1$ the hysteresis is present also at $T/T_c \approx 0.94$, but it manifests itself only through the presence of the plateau below $\mu_0 H_0 \approx 0.05$~T.

A detailed investigation of $R_s(H_0)$ of the samples with $x=0.1$ and $x=0.3$ at temperatures near $T_c$ has shown that for $x=0.1$ the hysteresis is detectable up to $T/T_c \approx 0.96$; moreover, for $T/T_c \gtrsim 0.87$ it manifests itself only through the presence of the plateau. For $x=0.3$, the hysteresis disappears at $T/T_c \approx 0.86$. We have also measured the field-induced variations of $R_s$ in a sample with $x=0.2$; results not here reported have shown a magnetic behavior similar to that obtained for $x=0.1$, i.e. plateau in the decreasing-field branch of the $R_s(H_0)$ curve and hysteresis visible up to temperatures close to $T_c$.

The results obtained in the sample with $x=0.4$ can be fully justified for any temperature using the same model we have used for $T/T_c=0.2$ (Fig.~\ref{fig1}c). However, near $T_c$ in order to fit the data it is necessary to use a distribution function of $T_c$, increasing the number of fitting parameters. For this reason, we have not reported in Fig.~\ref{nearTc}c the best-fit curve.

In Ref.~\cite{noiGIANNI} we have investigated $R_s(H_0)$ in a two-gap \mgb\ sample in the range of fields at which the superconductivity coming from the $\pi$ band is almost suppressed. In this region ($H_0 \geq H_{c2}^{\pi}$), one expects that the flux lines assume a "conventional" structure and that all the charge carriers coming from the $\pi$ band are quasiparticles. On this hypothesis, we have elaborated a model that quantitatively describes the field dependence of $R_s$ for $H_0 \geq H_{c2}^{\pi}$. By fitting the data, we have shown that, at least in the temperature range in which the hysteresis manifests itself only through the presence of the plateau, the values of $H_{c2}^{\pi}(T)$ coincide with the values of $H_0$ at which the plateau in the decreasing-field branch of $R_s(H_0)$ starts. The result we obtained in ~\cite{noiGIANNI} allows to determine the temperature dependence of $H_{c2}^{\pi}$ at temperatures near $T_c$. By applying the same arguments of Ref.~\cite{noiGIANNI} to the results obtained in the sample with $x=0.1$, we have deduced a linear dependence of $H_{c2}^{\pi}(T)$ for $T\geq 29$~K, which is well described by the law $H_{c2}^{\pi}(T)=(0.57-0.017~T)$~T.

\section{Conclusion}
We have measured the magnetic-field-induced variations of the mw surface resistance in ceramic samples of Mg$_{1-x}$(LiAl)$_x$B$_2$, with $x=0.1 \div 0.4$. At low temperatures, we have detected a magnetic hysteresis in all the investigated samples, but only that observed in the sample with $x=0.4$ can be justified in the framework of the standard theory of fluxon dynamics, provided that one considers the distribution of the magnetic induction inside the sample due to the critical state of the fluxon lattice. In the other samples, the decreasing-field branch of the $R_s(H_0)$ curve exhibits a plateau from $H_0$ much larger than the fluxon-first-penetration field down to zero, which cannot be explained in the framework of critical-state models. It has been already shown that this characteristic of the decreasing-field branch of the $R_s(H_0)$ curve is peculiar of the two-gap \mgb\ samples and confirms that in the samples with $x<0.4$ the two gaps have distinct values. On the contrary, the fact that the sample with $x=0.4$ have shown a $R_s(H_0)$ curve compatible with the results expected from standard models for fluxon dynamics is consistent with the merging of the two superconducting gaps into a single value, observed in this sample by point-contact Andreev-reflection measurements. Our results confirm that by measuring the field-induced variations of the  mw resistance, one may discriminate the gap structure of \mgb\ samples.

\section*{Acknowledgements}
The authors are very glad to thank M. Putti for her interest to this work and G. Napoli for technical assistance.


\end{document}